\documentclass[a4paper,10pt]{amsart}

\usepackage{amssymb}
\usepackage{graphicx}
\usepackage[latin1]{inputenc}
\usepackage{latexsym}
\usepackage{url}
\newcommand{\R}{\mathbb{R}}
\newcommand{\set}[1]{\{#1\}}
\renewcommand{\vec}[1]{\textbf{#1}}
\newcommand{\vun}{\vec{1}}

\newcommand{\sumi}{\sum_{i,\: i\rightarrow j}}
\newcommand{\sumj}{\sum_{j,\: i\rightarrow j}}

\newcommand{\sumk}{\sum_{k,\: k\rightarrow j}}
\renewcommand{\b}[1]{\textbf{#1}}
\newcommand{\grad}{\textnormal{grad }}

  \newtheorem{prop}{Proposition}
  \newtheorem{thm}[prop]{Theorem}

\begin{document}

\title{Iterative Filtering\\ for a Dynamical Reputation System}
\author{Cristobald de~Kerchove \and Paul Van~Dooren}
\maketitle

\begin{abstract}
The paper introduces a novel iterative method that assigns a
reputation to $n+m$ items: $n$ raters and $m$ objects. Each rater
evaluates a subset of objects leading to a $n \times m$ rating
matrix with a certain sparsity pattern. From this rating matrix we
give a nonlinear formula to define the reputation of raters and
objects. We also provide an iterative algorithm that superlinearly
converges to the unique vector of reputations and this for any
rating matrix. In contrast to classical outliers detection, no
evaluation is discarded in this method but each one is taken into
account with different weights for the reputation of the objects.
The complexity of one iteration step is linear in the number of
evaluations, making our algorithm efficient for large data set.
Experiments show good robustness of the reputation of the objects
against cheaters and spammers and good detection properties of
cheaters and spammers.
\end{abstract}

\section{Introduction}

There is an important growth of sites on the World Wide Web where
users play a crucial role: they provide trust ratings to objects
or even to other raters. Such sites may be commercial, where
buyers evaluate sellers or articles (Ebay, Amazone, etc.), or they
may be opinion sites, where users evaluate objects (Epinions,
Tailrank, MovieLens, etc.). But websites are not the only place
where we can find ratings between users and items: the simple fact
to link to another webpage is considered by search engines as a
positive evaluation (Google, Yahoo, etc.). Therefore the good
working of auction systems, opinion websites, search engines, etc.
depends directly on the reliability of their raters and on the
treatment of all the data. Trust and reputation in the electronic
market gives a necessary transparency to their users. For example,
in 1970, Akerloff~\cite{Akerloff} pointed out the information
asymmetry between the buyers and the sellers in the market for
lemons. The former had more information than the latter, making
hard trusting trading relationships. From what precedes, two
questions naturally arises:\\
\\
\textit{- What should be the reputation of the evaluated items?}\\
\textit{- How can we measure the reliability
of the raters?}\\
\\
We will distinguish the \emph{reputation}, that is what is
generally said or believed about a person's or thing's character
or standing, and the \emph{reliability}, that is the subjective
probability by which one expects that a rater gives an evaluation
on which its welfare depends. Let us remark that many technics
only calculate the reputations of items. Sometimes reputation and
reliability have the same value as it is the case in eigenvector
based technic where the reputation of any individual depends on
the reputations of his raters~\cite{Google,Eigentrust}. In these
methods, they construct a stochastic matrix from the network and
the ratings, then the eigenvector of that matrix gives the
reputations. Another part of the literature concerns the
propagation of trust (and
distrust)~\cite{Kumar,Mui,Richardson,Baras} where they define
trust metrics between pairs of individuals $(A,B)$ looking at the
possible paths linking $A$ with $B$. So reputations depend on the
point of view of the user and these methods differ from ours that
assigns one global reputation for each item.\\

Our method weights the evaluations of the raters. A small weight
is a natural way to tackle the problem of attackers in reputation
systems. Therefore, the method gives two values for a user: his
reputation depending on his received evaluations and his weight
that influence the impact of his given evaluations. The algorithm
is based on an iterative refinement that is guarantied to converge
to a reputation score and a reliability score for each item: at
each step the reliability of a rater is calculated according to
some distance between his given evaluations and the reputations of
the items he evaluates. This distance is interpreted as the
\emph{belief divergence}. Typically, a rater diverging to much
from the group will be distrusted after convergence. The same
definition of distance appears in \cite{It_Filt,C_F_log,Jiminy}
and is used for the same issue. In \cite{It_Filt}, the function
that determines the weights is different. This difference makes
their algorithm sensitive to initial conditions without any
guaranty of convergence. Moreover, they are in the less general
case where it is supposed that every rater evaluates all items. In
\cite{C_F_log}, they want to tackle the problem of spammers in
collaborative filtering where previous evaluations are used to
predict future evaluations. Again the same definition of distance
allows to penalize the divergent raters. Even though the function
that assigns the weights for the evaluations is the same, there is
no iterative procedure but only a simple step is applied. In
\cite{Jiminy}, they use another function to determine the weights:
the log-likelihoods, but again only a simple step is applied. We
show in section \ref{sec_exp} the advantage to apply more than one
step in the iterative filtering. Indeed, each step separates a
little more the outliers.\\

Let us remark that beside the refinement process of the
reputations and the outlier detection given by our procedure,
other applications can take advantage of these data. For example,
\cite{C_F_log} want to remove spammers to improve collaborative
filtering. Similarly in \cite{Donovan}, they propose a framework
to take into account the different qualities of ratings for
collaborative filtering. Hence they weight each rating according
to its reliability, these weights can be those obtained by the
iterative filtering we described.\\

In the sequel, we first explain in section \ref{sec_method} how
the reputation vector for the objects and the weights for the
evaluation are built. Moreover, we develop the algorithm
\verb"Reputation" that calculates these values, and we explain its
interpretation and its properties of convergence. Then in section
\ref{sec_exp}, our experiments test the robustness of our method
against attackers and show several iterations on graphics. Finally
in section \ref{sec_conc}, we point out possible extensions and
experiments for our method.

\newpage
\section{The iterative method}\label{sec_method}

Before to develop the model and the algorithm, we introduce the
main notations in the following tabular.\\
\\
\begin{tabular}{|r||l|}
  \hline \label{tab_not}
  \: Notations \:    & \quad Definitions \\
  \hline
  $n$, $m$, $m_i$ \: & \: \# of \emph{raters},\# of \emph{items},\\&\: \# of items evaluated by $i$.\\
  \cline{2-2}
  $E$ \: & \: The $n\times m$ \emph{rating matrix}:\\
  & \: $E_{ij}$ is the evaluation given\\&\: by rater $i$ to item $j$\\
  \cline{2-2}
  $A$ \: & \: The $n\times m$ \emph{adjacency matrix}:\\
  & \: $A_{ij}=1$ if rater $i$ evaluates $j$,\\&\: otherwise $A_{ij}=0$ (and $E_{ij}=0$)\textnormal{ } \\
  \cline{2-2}
  $T$ \: &\:  The $n\times m$ \emph{trust matrix} of evaluations\\
  $t$ \: &\:  The $n\times 1$ \emph{trust vector} of raters\\
  $r$ \: &\: The $m\times 1$ \emph{reputation vector} of items\\
  $\vun$ \: & \: The $n\times 1$ or $m\times 1$ vector of ones\\
  \cline{2-2}
  $\sumi$ \: &\: Sum over the set $\set{i | A_{ij}=1}$ \\
  $\sumj$ \: & \: Sum over the set $\set{j | A_{ij}=1}$ \\
  \hline
\end{tabular}\\
\\
Without loss of generality, we will consider ratings in the
interval $[0,1]$, i.e. $E\in [0,1]^{n\times m}$ and therefore the
reputation vector $r$ will belong to $[0,1]^m$. Moreover, the
trust matrix $T$ and the trust vector $t$ are nonnegative, i.e.
the entries of $T$ and $t$ are nonnegative.

\subsection{The model}\label{sec_model}

As already said in the introduction, the reputations of the items
essentially depend on the evaluations they receive. These latter
are weighted according to their reliability. In that way, the
reputation of item $j\in\set{1,\dots,m}$ is obtained by taking the
weighted sum of its evaluations, i.e.
\begin{equation}\label{eq_r}
r_j = \sumi W_{ij} E_{ij},\quad \sumi W_{ij}=1
\end{equation}
And we define the matrix $W$ from the trust matrix $T$ in the
following way:
\begin{equation}\label{eq_Wij_Tij}
W_{ij} = \frac{T_{ij}}{\sumk T_{kj}},\quad i=1,\dots,n,\quad
j=1\dots,m.
\end{equation}
In that manner, evaluations with a higher trust value are taken
into account more for the reputation vector. Now the important
role is played by the trust matrix $T$, its definition is given in the next section.\\

\subsection{The trust matrix}

Let us describe the trust matrix that assigns a measure of
confidence to each rating. The inputs of the trust matrix are the
rating matrix $E$ and the reputation vector $r$. Formally, we
define the belief divergence of rater $i\in\set{1,\dots,n}$ as the
estimated variance of the $i^{th}$ row of $E$:
\begin{equation}\label{eq_d}
d_i = \frac{1}{m_i}\sumj (E_{ij}-r_j)^2,
\end{equation}
where $m_i$ is the number of items evaluated by $i$. That
definition is somewhat similar to the one proposed
in~\cite{C_F_log} where $d$ is used to penalize those raters that
have an high belief divergence. The resulting trust matrix is
\begin{equation}\label{eq_Tij}
T_{ij} = c_j - d_i,
\end{equation}
for any evaluation from $i$ to $j$. The parameters $c_j$ are
chosen such that the entries of $T$ are nonnegative. Moreover
$c_j$ are discriminating in the sense that they influence the
ratios $T_{ij}/T_{kj}$ for $i,k\in\set{1,\dots,n}$ and
$j\in\set{1,\dots,m}$. Typically, the smaller $c_j$, the more
spammers evaluating object $j$ are penalized. In order to have a
trust value for each raters, we also define the trust vector $t$:
\begin{equation}\label{eq_t}
t_i = d_{max} - d_i,
\end{equation}
where $d_{max}$ is the maximum of the elements of the vector $d$.

\subsection{The algorithm}\label{sec_algo}

From equations (\ref{eq_r}-\ref{eq_t}), we can derive the
algorithm \texttt{[r t] = Reputation(E,A,c)} that takes as inputs
a rating matrix $E$, an adjacency matrix $A$ and a $m\times 1$
vector $c$ of parameters. Then it iteratively calculates the
reputation vectors $r$ and the trust matrix $T$. Eventually, the
algorithm gives the reputation and the trust vector. The
description is given in four steps corresponding to
the initialization, two updates for $r$ and $T$ and the calculation of the trust vector $t$.\\
\begin{itemize}
\item[\b{(i)}] Initialization of matrix $T$: every rater is evenly
trusted, i.e. $T_{ij}=1$ for $i=1,\dots,n$ and
$j=1,\dots,m$.\\

\item[\b{(ii)}] The matrix of weights and the reputation vector
are calculated from $T$. For $i=1,\dots,n$ and $j=1,\dots,m$:
$$
W_{ij} = \frac{T_{ij}}{\sumk T_{kj}},\quad\quad r_j = \sumi W_{ij}
E_{ij}.
$$

\item[\b{(iii)}] The belief divergence and the new trust matrix
are calculated from $r$. For $i=1,\dots,n$ and $j=1,\dots,m$:
$$
d_i = \frac{1}{m_i}\sumj (E_{ij}-r_j)^2,\quad\quad T_{ij} = c_j -
d_i.
$$
If the $i^{th}$ row of $T$ is zero, then replace it by a row of
ones.\\
Repeat steps \b{(ii)} and \b{(iii)} until convergence.\\

\item[\b{(iv)}] The trust vector $t$ is given by $\max_{k} d_k -
d_i$.
\end{itemize}
Let us remind that the input parameters $c$ used in step \b{(iii)}
are chosen sufficiently large such that the entries of $T$ are
nonnegative at each iteration.\\

\begin{figure}
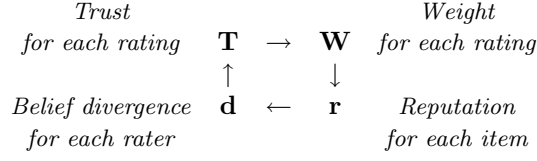

$$
\begin{array}{ccccc}
\small\textit{Trust}& &  &  &\small\textit{Weight}\\
\small\textit{for each rating}&\bf{T}&\rightarrow&\bf{W}&\small\textit{for each rating}\\
&\uparrow & & \downarrow&\\
\small\textit{Belief divergence}&\bf{d} & \leftarrow & \bf{r} &
\small\textit{Reputation}\\
\small\textit{for each rater}&&&&\small\textit{for each item}
\end{array}
$$
\caption{Cycle of one iteration of the
algorithm.}\label{fig_cycle}
\end{figure}

One iteration of the algorithm \texttt{Reputation} is schematized
in figure~\ref{fig_cycle}. A slight modification allows dynamical
evaluations. In that case, the rating matrix $E$ changes at each
iteration, i.e. $E[k]$ with $k=1,2,\cdots$. A direct way to update
$r$, $T$ and $t$, is given when the first step takes as initial
vector the previous trust matrix, and steps \b{(ii)} and \b{(iii)}
are not repeat until convergence, but a certain number of times.

\subsection{Interpretation of the solution}\label{sec_inter}

The algorithm in section \ref{sec_algo} converges to the unique
solution of equations (\ref{eq_r}-\ref{eq_Tij}), see next section.
Let $r^*$ represents that solution\footnote{the corresponding
matrices and vectors $W^*$, $T^*$, $d^*$ and $t^*$ follow from
$r^*$} and for the sake of simplicity, let parameters $c_j$ be
equals to a same constant $c_0$. Then $r^*$ is the maximizer of
the following scalar function: $\psi:[0,1]^m\rightarrow\R$ with
\begin{eqnarray}
\psi(r) &=& \sum_{i=1}^{n} \sumj  T_{ij}^2 \\
        &=& \sum_{i=1}^{n} m_i \left( c_0 - \frac{1}{m_i} \sumj
        (E_{ij}-r_j)^2\right) ^2.\label{eq_psi_t}
\end{eqnarray}
It is indeed enough to observe that $\text{grad }\psi(r^*)=0$. In
other words, $r^*$ maximizes the Frobenius norm of the sparse
trust matrix $T$. Maximizing such a norm roughly means that some
total degree of confidence over the raters is maximized. More
formally, let assume that the entries $E_{ij}$ are i.i.d.$\sim
N(r_j,\sigma^2)$. In that case, the degree of confidence we can
have in evaluation $E_{ij}$ is given by
$$
\log Pr(E_{ij}|r_j) = \textnormal{cst} -
\frac{1}{2\sigma^2}(E_{ij}-r_j)^2,
$$
and by summing the evaluations of $i$ and choosing the appropriate
constant, we obtain the relation
$$
T_{ij} = \frac{2\sigma^2}{m_i} \sumk\log Pr(E_{ij}|r_j).
$$
In other words, the trust matrix is the normalized sum of the
degrees of confidence we have in the evaluations of the raters.
Then the maximizer of the Frobenius norm of $T$ is
$$
r^* = \arg\max_{r\in[0,1]^m} \sum_{i=1}^{n} \frac{1}{m_i} \left(
\sumj\log Pr(E_{ij}|r_j)\right)^2.
$$
Another writing of the function $\psi$ leads to
$$
\psi(r) = -2c_0\sum_{i=1}^{n} \sumj (E_{ij}-r_j)^2 +
\sum_{i=1}^{n} \frac{1}{m_i} \left( \sumj (E_{ij}-r_j)^2
\right)^2,
$$
and the maximizer of the first expression is simply the average of
the evaluations for each item $j$, i.e. $\sumi
E_{ij}/|i,\:i\rightarrow~j|$, and the maximizer of the second
expression necessarily belongs to the border of the hyper cube,
i.e. $\{0,1\}^m$. The solution $r^*$ is then a compromise between
both terms in which the parameters $c_0$ plays the role of a
weighting factor. For large $c$, the algorithm will give the
average of the evaluations for each item.

\subsection{Properties of convergence}\label{sec_conv}

\begin{figure}[t]
\centering
\includegraphics[width=6cm]{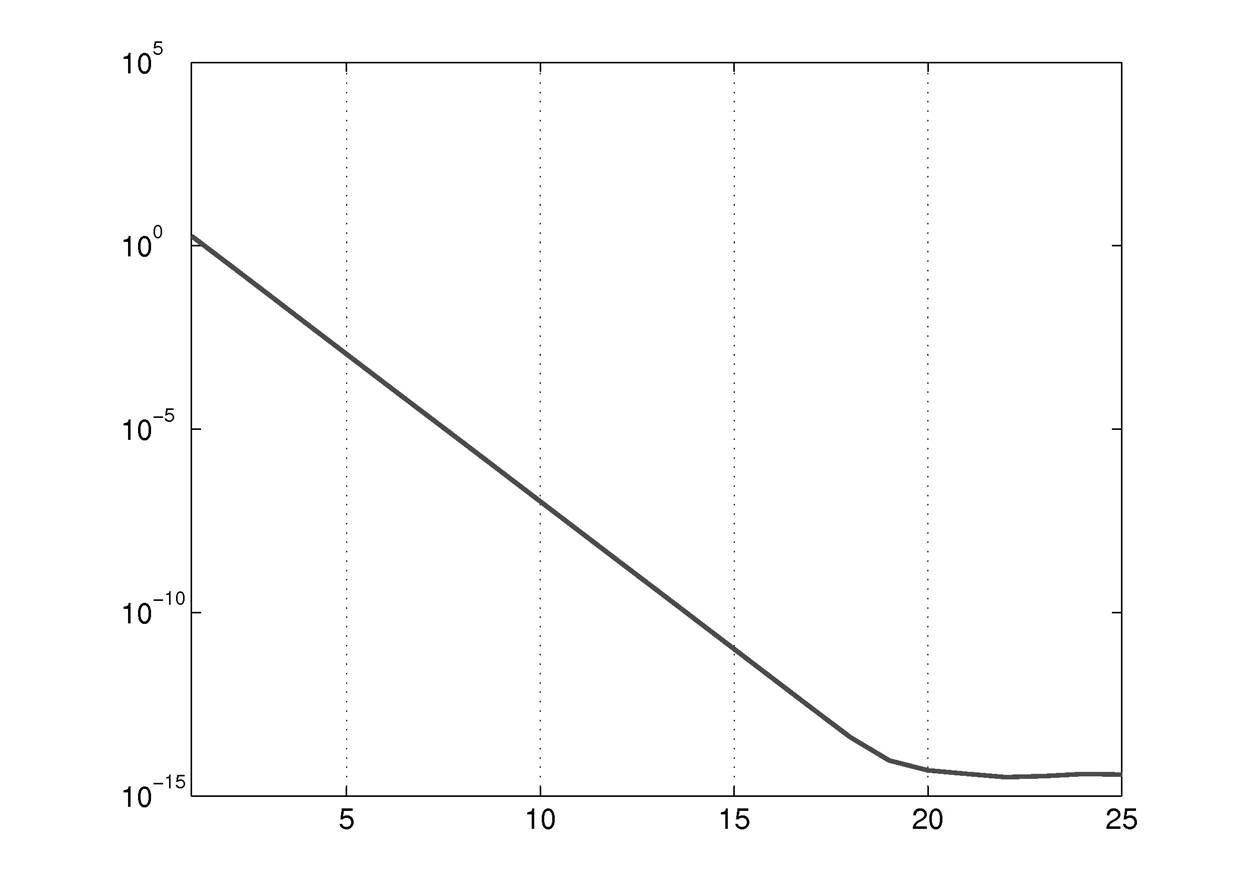}
\caption{X-Axis: \# of iterations. Y-Axis: the euclidian norm of
the error, i.e. $\|r[k]-r^*\|_2$. The graphic was obtained from a
$943\times 1682$ sparse rating matrix representing $10^5$
evaluations.}\label{fig_err}
\end{figure}
In this section we analyze the convergence of \texttt{Reputation}
given in section \ref{sec_algo} and its rate of convergence. For
the sake of clearness, we restrict ourselves to the main and
important steps in the proof avoiding the technical points.
\begin{thm}
For $c_j$ chosen such that the matrix $T$ is nonnegative at each
step, the iteration in the algorithm
\texttt{Reputation(E,A,c)}~converges to the unique vector $r^*$.
\end{thm}
\begin{proof}(Only a sketch).
For the sake of simplicity, we let parameters $c_j$ be equals to a
same constant $c_0$. We first prove the unicity of $r^*$ and then
we prove the convergence result.

It can be shown that the scalar function $\psi:[0,1]^m\rightarrow
\R:r\mapsto\psi(r)$ defined in \eqref{eq_psi_t} is continuous and
quasiconcave. It has therefore a unique maximizer on $[0,1]^m$
that corresponds to the vector $r^*$.

Let $r_k$ and $r_{k+1}$ be two successive iterations of $r$. It
can be shown by simple developments that these two vectors are
linked by the following expression:
\begin{equation}
r_{k+1} = r_k + \alpha(r_k)\cdot\grad \psi(r_k),
\end{equation}
where $\alpha(r_k)\geq (4c_0)^{-1}>0$ and $\grad \psi(r_k)$ is the
gradient of $\psi$ in $r_k$ pointing to the direction of greatest
ascent. Therefore one iteration corresponds to take the direction
of greatest ascent and make a step of length $\ell_k =
\alpha(r_k)\cdot\| \grad \psi(r_k)\|_2$. Moreover, it can be shown
that $\ell_k$ is such that we have strict ascent:
$$
\psi(r[k]) < \psi(r[k+1]).
$$
Finally, $\ell_k$ is also lower bounded by $(4c_0)^{-1}\| \grad
\psi(r[k])\|_2$ and therefore the iteration on $r$ monotonically
converges to the maximizer of $\psi$ on $[0,1]^m$.
\end{proof}
Numerical experiments show a linear rate of convergence for the
vector $r$. As shown in Figure \ref{fig_err}, the logarithm of the
error decreases linearly and stabilizes after $20$ steps. It is
possible to speed up the rate of convergence by using a Newton
method provided that we are close enough to $r^*$. Then the rate
of convergence becomes quadratic making our algorithm efficient
for large data set.
\begin{figure}[t!]
\includegraphics[width=8cm]{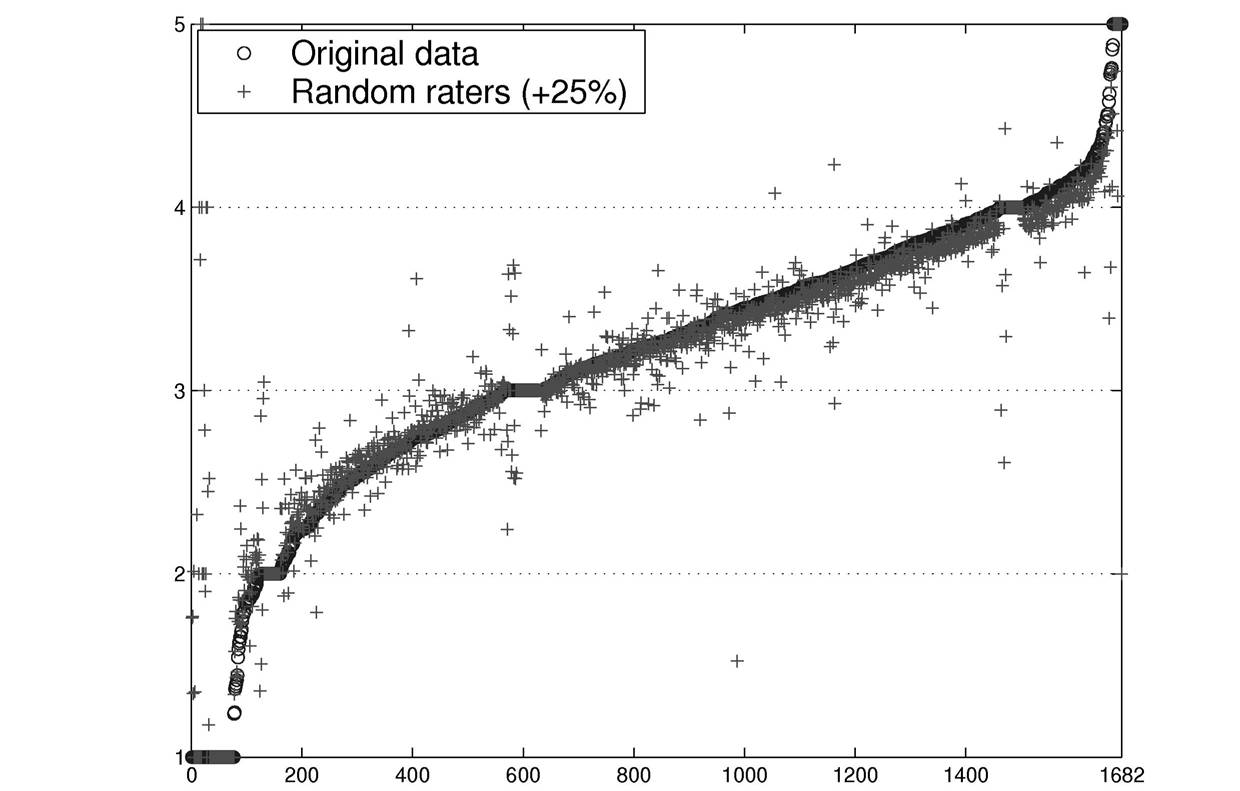}
\includegraphics[width=8cm]{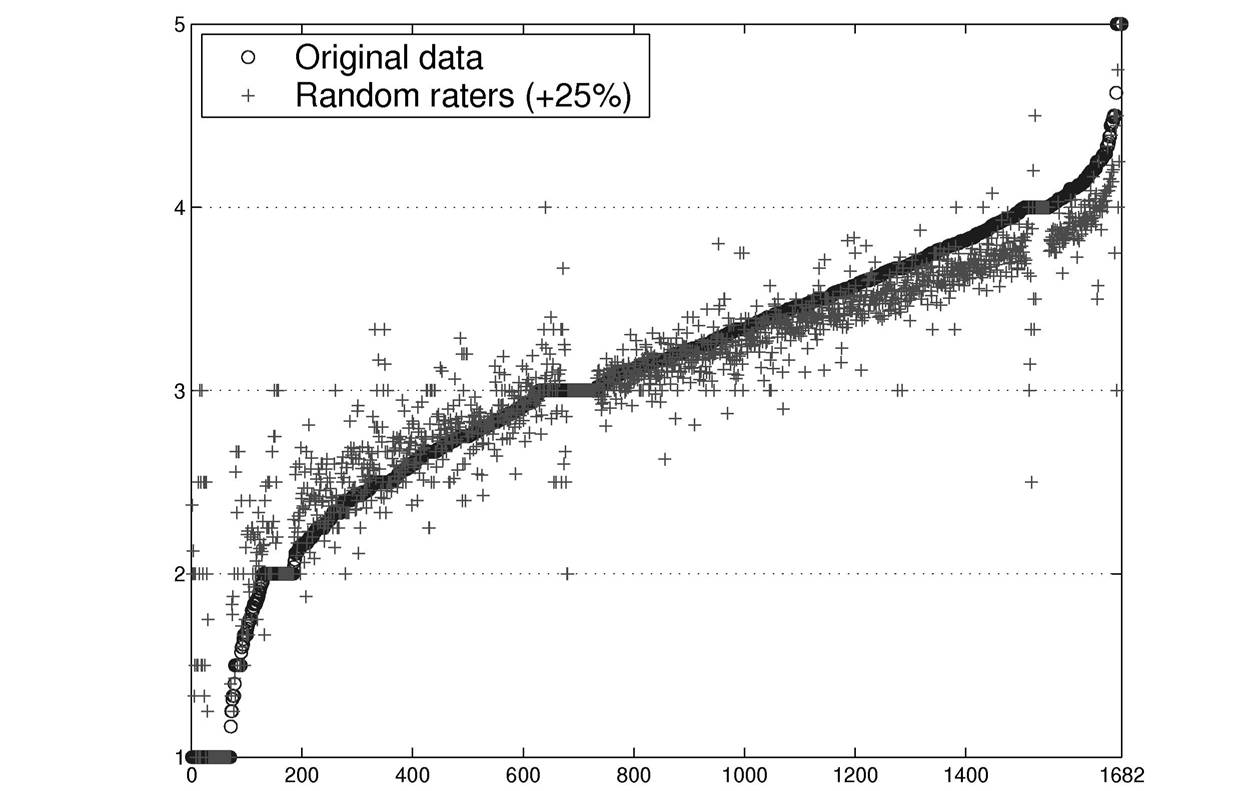}
\caption{X-Axis: the sorted movies according to their reputations
before the addition of random raters. Y-Axis: their reputations
according to our algorithm (Top) and to the average
(Bottom).}\label{fig_rand}
\end{figure}

\section{Experiments}\label{sec_exp}

Our experiment concerns a data set\footnote{The MovieLens data set
used in this paper was supplied by the GroupLens Research
Project.} of 100,000 evaluations given by 943 users on 1682 movies
and raging from $1$ to $5$. Each user has rated at least 20
movies.

In order to simulate the robustness of the algorithm
\verb"Reputation", two types of behavior are analyzed in the
sequel: first, raters that give random evaluations, and second,
spammers that try to improve the reputation of their preferred
item.

\subsection{Robustness against random raters}

We added to the original data set $237$ raters evaluating randomly
some items. In that manner, $20\%$ of the raters give random
evaluations. Let $r^*$ and $\tilde{r}^*$ be respectively the
reputation vector before and after the addition of the random
raters. If the reputation vector is calculated according to
\verb"Reputation", then the $1$-norm difference between $r^*$ and
$\tilde{r}^*$ is
$$
\|r^*-\tilde{r}^*\|_1=182,
$$
if the reputation vector is the average of the evaluations for
each item, then the $1$-norm difference between $r^*$ and
$\tilde{r}^*$ increases:
$$
\|r^*-\tilde{r}^*\|_1=259.
$$
Figure \ref{fig_rand} illustrates this perturbation due to the
addition of random raters.
\begin{figure}[t!]
\centering
\includegraphics[width=4cm]{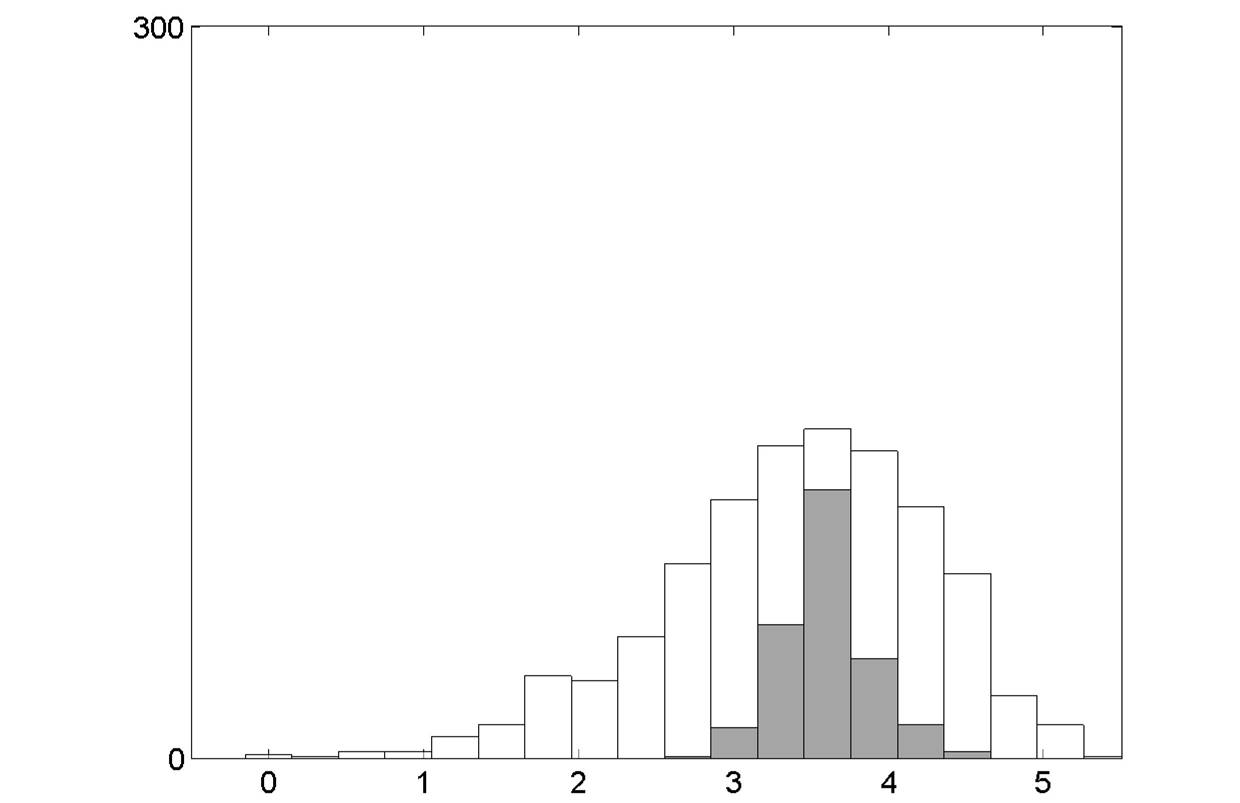}\\
\includegraphics[width=4cm]{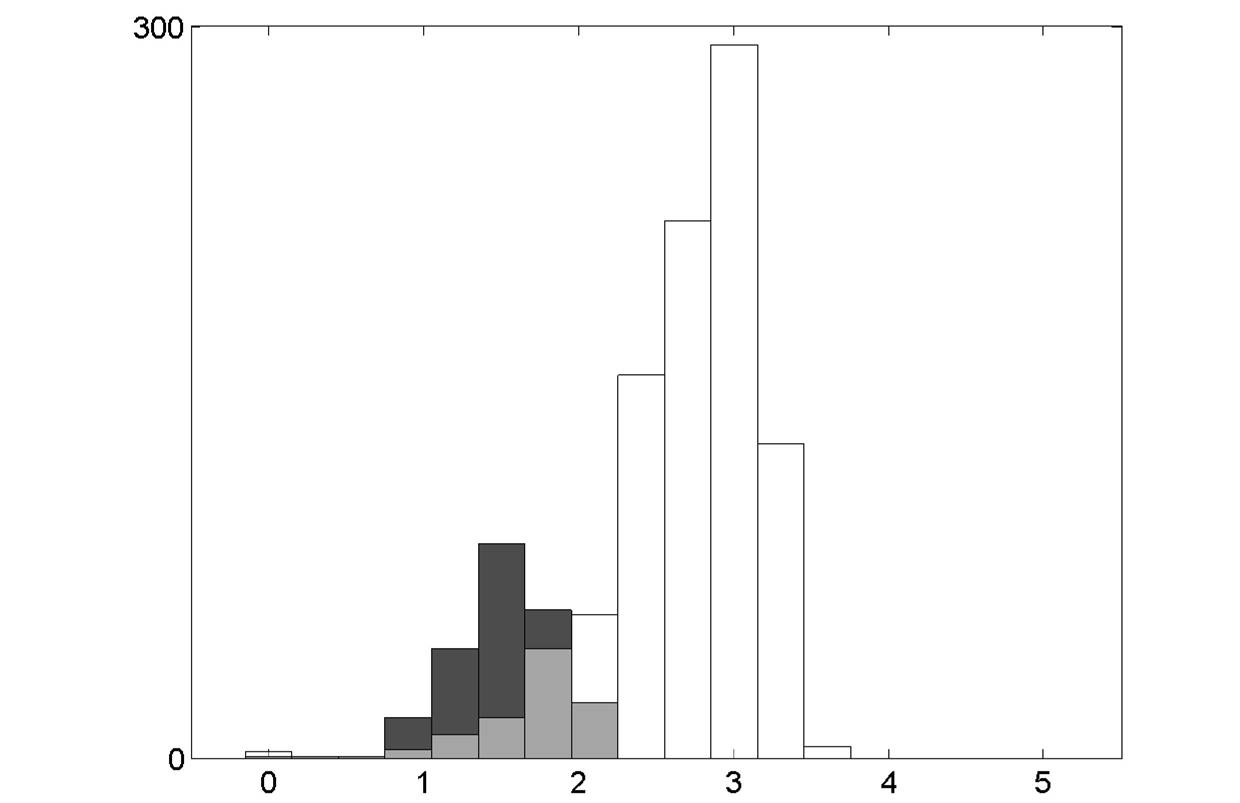}
\includegraphics[width=4cm]{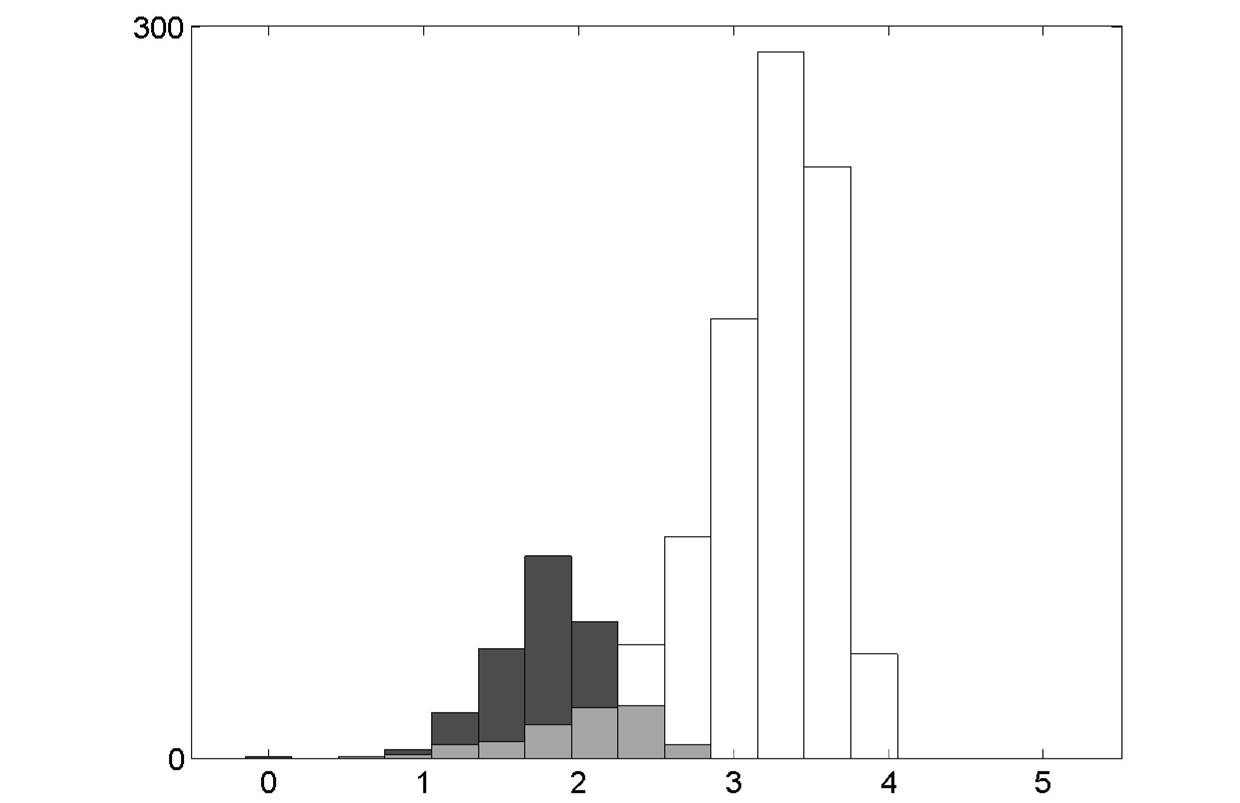}
\caption{X-Axis: the trust values for the raters. Y-Axis: the
density after one iteration (Top), after two iterations (Left),
and after convergence (Right). In black: the random raters. In
white: the original raters. In grey: both
raters.}\label{fig_dist_rand}
\end{figure}
The reputations are better preserved when using \verb"Reputation".
It turns out that the reputations given by \verb"Reputation" take
less into account the random users. Moreover, one iteration of the
algorithm gives poor information to trust the raters, it is indeed
useful to wait until convergence, as seen in Figure
\ref{fig_dist_rand}.

\subsection{Robustness against spammers}

\begin{figure}[t!]
\includegraphics[width=8cm]{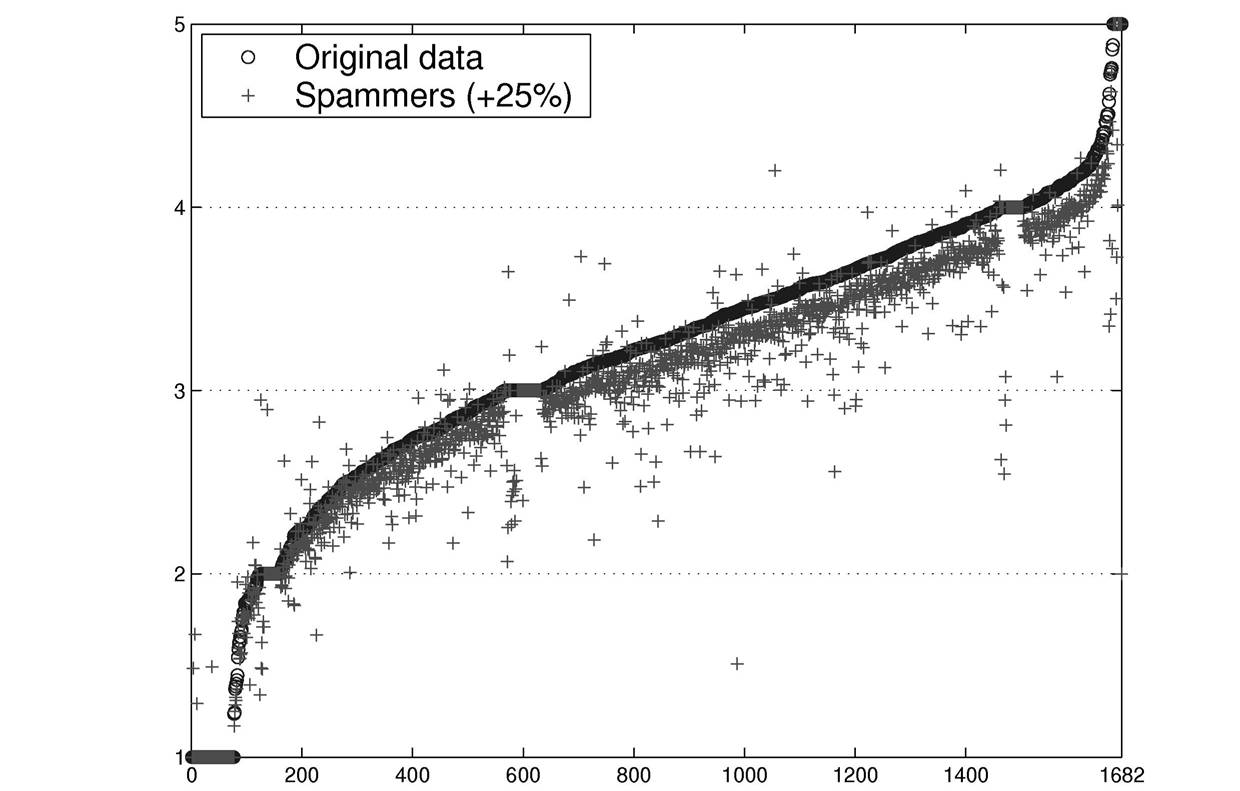}
\includegraphics[width=8cm]{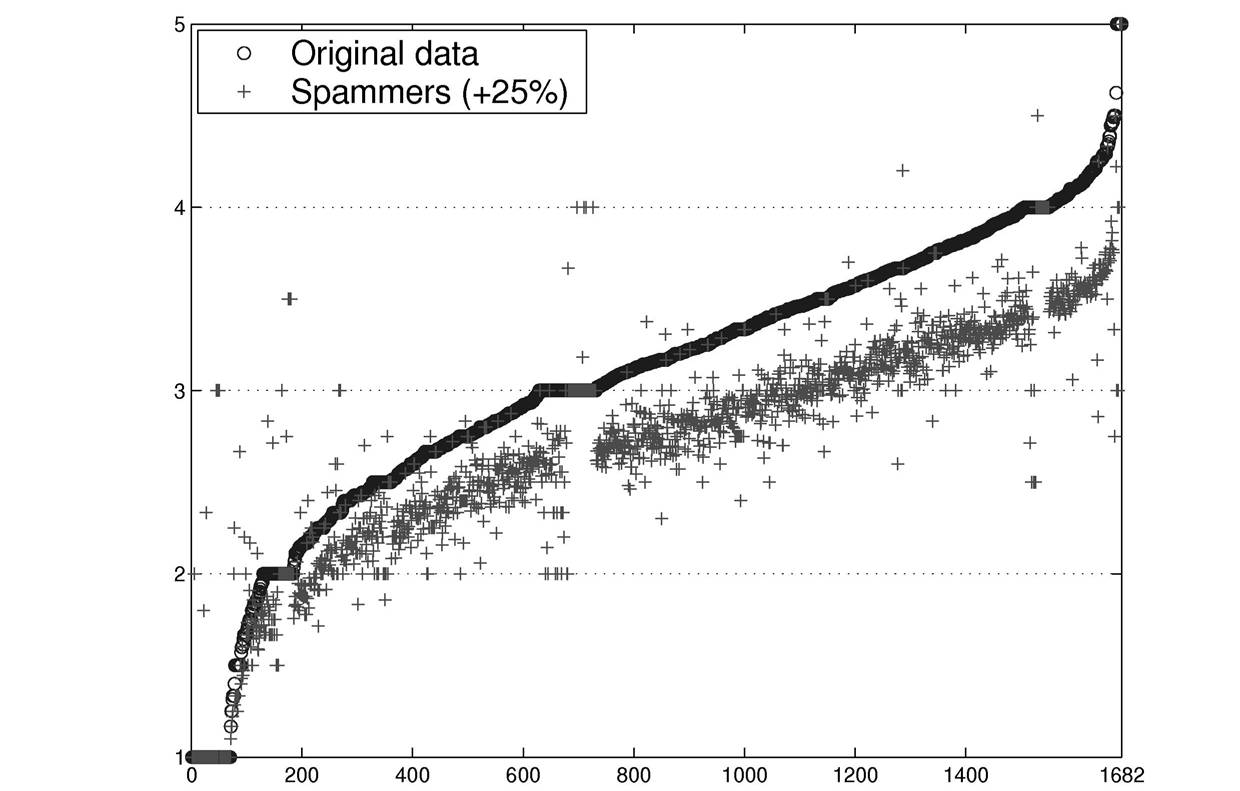}
\caption{X-Axis: the sorted movies according to their reputations
before the addition of spammers. Y-Axis: their reputations
according to our algorithm (Top) and to the average
(Bottom).}\label{fig_spam}
\end{figure}

\begin{figure}[t]
\centering
\includegraphics[width=4cm]{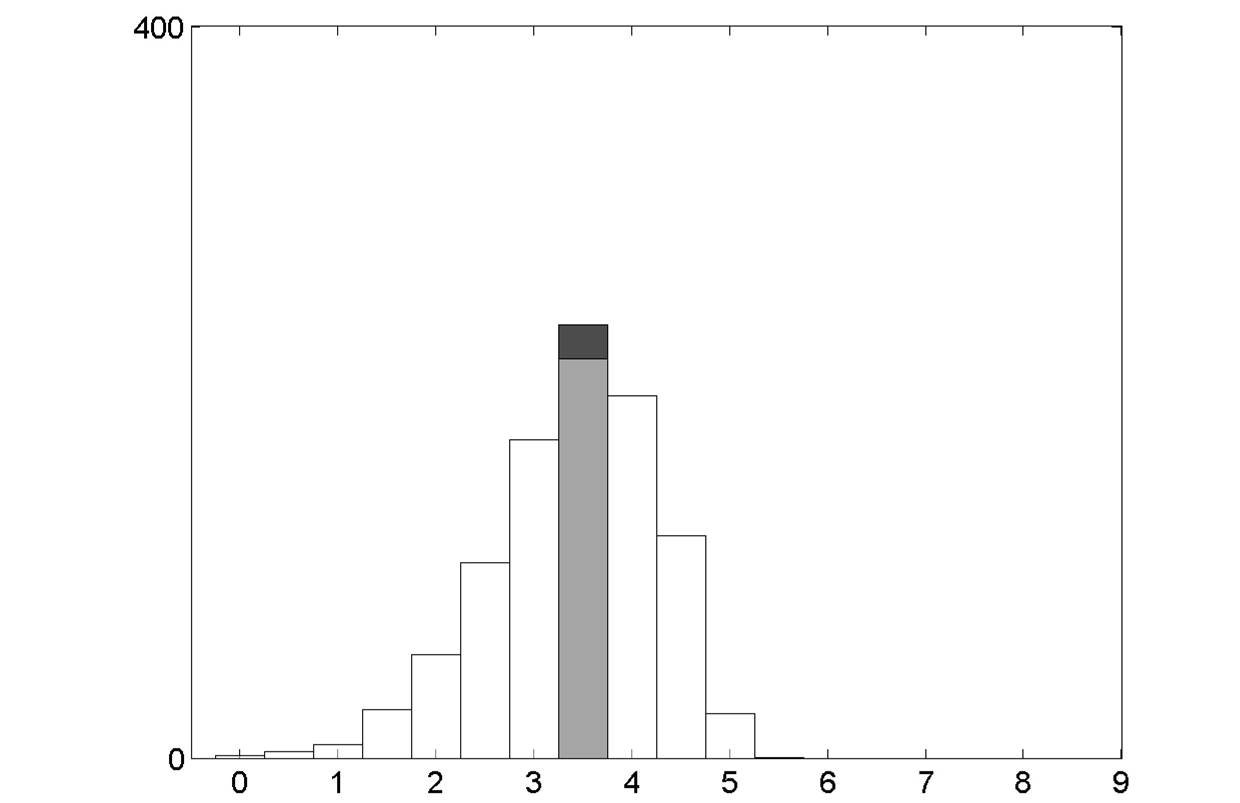}\\
\includegraphics[width=4cm]{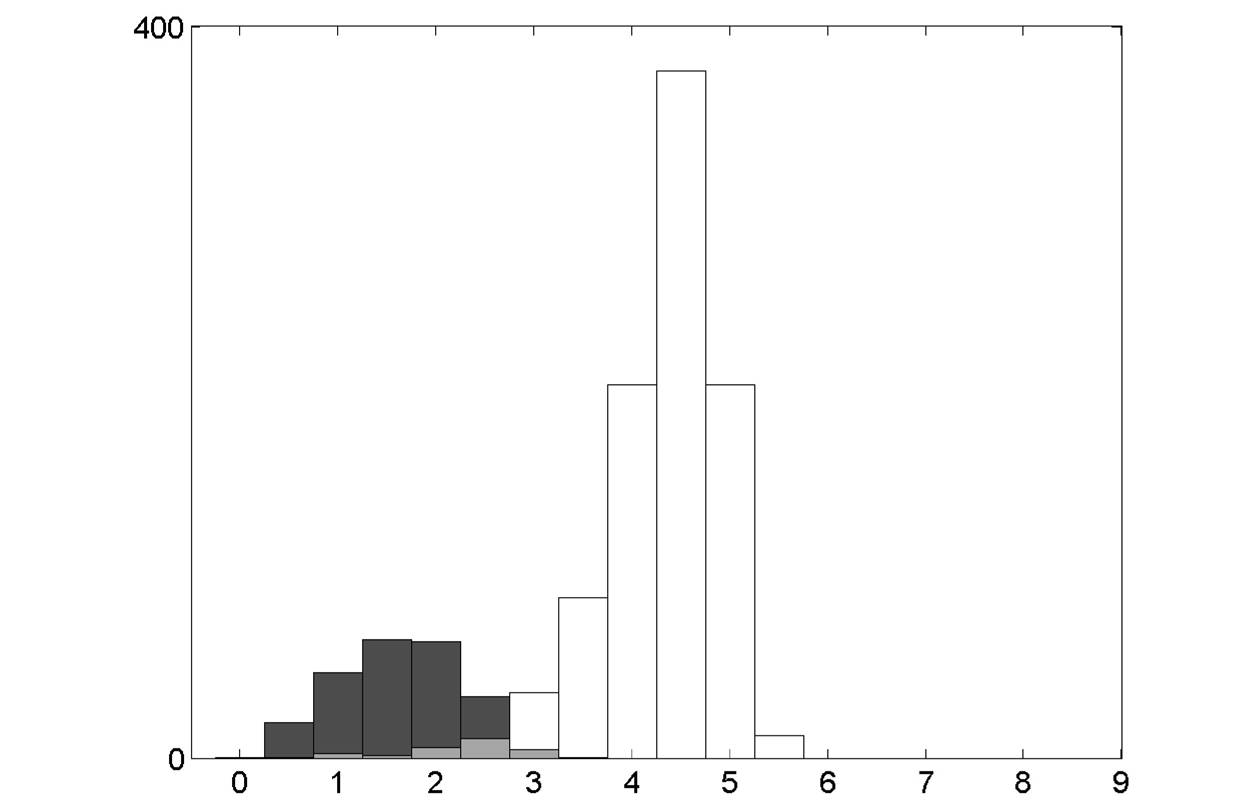}
\includegraphics[width=4cm]{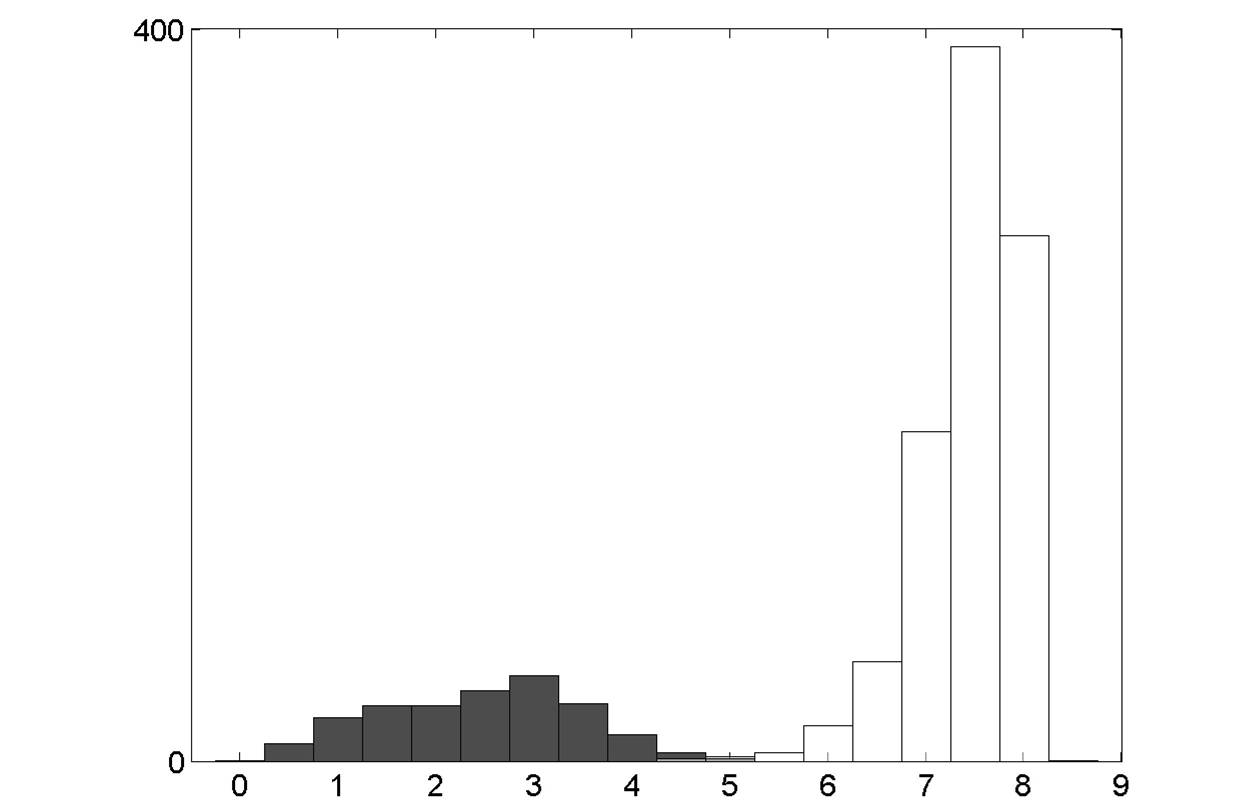}
\caption{X-Axis: the trust values for the raters. Y-Axis: the
density after one iteration (Top), after two iterations (Left),
and after convergence (Right). In black: the spammers. In white:
the original raters. In grey: both users.}\label{fig_dist_spam}
\end{figure}

We now added to the original data set $237$ spammers giving always
$1$ except for their preferred movie, which they rated $5$. Let
$r^*$ and $\tilde{r}^*$ be respectively the reputation vector
before and after the addition of the random raters. If the
reputation vector is calculated according to \verb"Reputation",
then the $1$-norm difference between $r^*$ and $\tilde{r}^*$ is
$$
\|r^*-\tilde{r}^*\|_1=267,
$$
if the reputation vector is the average of the evaluations for
each item, then the $1$-norm difference between $r^*$ and
$\tilde{r}^*$ increases:
$$
\|r^*-\tilde{r}^*\|_1=638.
$$
Figure \ref{fig_spam} illustrates this perturbation due to the
addition of spammers.
The reputations are again better preserved when using
\verb"Reputation". Again the reputations given by
\verb"Reputation" take less into account the spammers. As
previously, one iteration of the algorithm gives poor information
to trust the raters, it is indeed useful to wait until
convergence, as seen in Figure \ref{fig_dist_spam}.
\section{Conclusion and Future Work}\label{sec_conc}

Our method described in the paper allows us to efficiently refine
reputations for evaluated objects from structured data. It is
based on the trust we can have in the evaluations of the raters,
and also in the raters themselves. The parameters $c_j$,
introduced in equation \ref{eq_Tij}, make the method flexible,
ranging from the average method, i.e. every rater is evenly
trusted, until the discriminating method that takes $c_j$ as small
as possible.

The experiments show interesting results of robustness even though
the behavior of the added outliers is somewhat naive. The weights
of spammers and random raters are low for the aggregation of the
reputation vector. However, other behaviors could be analyzed. For
example, clumsy raters could evaluate once correctly and once
randomly or we can imagine a more complicated mix of behaviors.
Typically, the weights of such raters will be between those of
spammers and those of honest raters. Last but not least, the
creative cheaters can use engineering to understand the working of
the system. The way to proceed is simple: they need to evaluate
correctly a group of item and then with that trust, they can rate
some target items. In order to significantly change the reputation
of these target items, they must have a number of coordinated
evaluations larger than the one of honest raters. Therefore such
cheaters can easily be disqualified by looking after coordinated
ratings to one or several items.

As said at the end of section \ref{sec_model}, the trust matrix
$T$ is the important point for the model. We define it by
$$
T_{ij} = c_j - d_i,
$$
for any evaluation from $i$ to $j$. Hence, the trust we have in
evaluation $T_{ij}$ decreases when the belief divergence $d_i$
increases. Other decreasing functions with respect to $d$ make
sense. For instance, $T_{ij}=e^{-c_j\:d_i}$ and
$T_{ij}=(c_j+d_i)^{-1}$ may perform well on some data sets. The
second definition with $c_j=0$ gives the method described in
\cite{It_Filt}. However, the main difference with our definition
lies in the uniqueness of the solution. It turns out that the
method in \cite{It_Filt} may have several solutions. On the other
hand, these solutions can be of interest if they reflect for
example two opinion trends.

In section \ref{sec_inter}, the solution is interpreted as the
maximizer of the Frobenius norm of $T$. It is possible to maximize
other norms of $T$. Then there can be several maximizers and these
maximizers will no more satisfy equation \eqref{eq_r}, but a
different one.

We see that our method can be extended towards different
directions. Our future work will address the interpretation and
the convergence of these extensions.

\subsubsection*{Acknowledgments.} This paper presents research results of the Belgian
Network DYSCO (Dynamical Systems, Control, and Optimization),
funded by the Interuniversity Attraction Poles Programme,
initiated by the Belgian State, Science Policy Office and
supported by the Concerted Research Action (ARC) "Large Graphs and
Networks" of the French Community of Belgium. The scientific
responsibility rests with its authors.

\end{document}